\documentclass[lettersize,journal]{IEEEtran}
\usepackage{url}
\usepackage{cite}
\usepackage{array}
\usepackage{textcomp}
\usepackage{stfloats}
\usepackage{verbatim}
\usepackage{graphicx}
\usepackage{algorithm}
\usepackage{algorithmic}
\usepackage{amsmath,amsfonts}
\usepackage[caption=false,font=normalsize,labelfont=sf,textfont=sf]{subfig}

\usepackage[printonlyused]{acronym}
\newacro{los} [LOS] {line-of-sight}
\newacro{em} [EM] {electromagnetics}
\newacro{ml} [ML] {maximum likelihood}
\newacro{mse} [MSE] {mean square error}
\newacro{snr} [SNR] {signal-to-noise ratio}
\newacro{rmse} [RMSE] {root mean square error}
\newacro{crlb} [CRLB] {Cram\'er-Rao lower bound}
\newacro{kld} [KLD] {Kullback–Leibler divergence}
\newacro{siso} [SISO] {single-input-single-output}
\newacro{mimo} [MIMO] {multiple-input multiple-output}
\newacro{mcrb} [MCRB] {misspecified Cram\'er-Rao bound}
\newacro{ris} [RIS] {reconfigurable intelligent surface}
\usepackage{color} 
\newcommand{\red}[1]{{\color{red}{#1}}} 
\newcommand{\blue}[1]{{\color{blue}{#1}}}

\newcommand{\magenta}[1]{{\color{magenta}{#1}}}
\newtheorem{remark}{Remark}

\usepackage{tikz} 
\usepackage{pgfplots} 
\usepackage[utf8]{inputenc}
\usepackage{circuitikz} 
\colorlet{Icol}{blue!55}
\ctikzset{bipoles/thickness=1.3}
\usepackage{hyperref}
\hypersetup{urlcolor=black}
\usepackage{diagbox}
\usepackage{makecell}
\usepackage{tabu,longtable}
\usepackage{threeparttable}
\usepackage{bm} % for bold symbol
\newcommand{\TT}{\mathsf{T}}

\newcommand{\pv}{{\bf p}}

\newcommand{\rv}{{\bf r}}

\newcommand{\xv}{{\bf x}}
\newcommand{\yv}{{\bf y}}
\newcommand{\zv}{{\bf z}}

\newcommand{\Am}{{\bf A}}
\newcommand{\Bm}{{\bf B}}

\newcommand{\Dm}{{\bf D}}

\newcommand{\Zm}{{\bf Z}}

\newcommand{\It}{{\rm I}}

\newcommand{\Rt}{{\rm R}}
\newcommand{\St}{{\rm S}}

\newcommand{\omegav}{\hbox{\boldmath$\omega$}}

\newcommand{\Phim}{\hbox{\boldmath$\Phi$}}

\usepackage{units} % use \unit command
\usepackage{amssymb} % for AMS symbols
\begin{document}
%\bstctlcite{IEEEexample:BSTcontrol}
% ------------------------------------------------------------------------------
\title{On the Impact of Mutual Coupling on \\RIS-Assisted Channel Estimation}

\author{Pinjun~Zheng,~\IEEEmembership{Student Member,~IEEE},
Xiuxiu~Ma,~\IEEEmembership{Student Member,~IEEE},\\
and~Tareq~Y.~Al-Naffouri,~\IEEEmembership{Senior Member,~IEEE}

\thanks{The authors are with the Electrical and Computer Engineering Program, Division of Computer, Electrical and Mathematical Sciences and Engineering (CEMSE), King Abdullah University of Science and Technology (KAUST), Thuwal, 23955-6900, Kingdom of Saudi Arabia. Email: \{pinjun.zheng; xiuxiu.ma; tareq.alnaffouri\}@kaust.edu.sa.}
\thanks{This publication is based upon the work supported by the King Abdullah University of Science and Technology (KAUST) Office of Sponsored Research (OSR) under Award No. ORA-CRG2021-4695.}
}

%\markboth{draft}{draft}
%\IEEEpubid{0000--0000/00\$00.00~\copyright~2021 IEEE}
\maketitle
% ------------------------------------------------------------------------------

\begin{abstract}
Amid the demand for densely integrated elements in techniques such as holographic reconfigurable intelligent surfaces (RISs), the mutual coupling effect has gained prominence.
By performing a misspecified Cram\'er-Rao bound analysis within an electromagnetics-compliant communication model, this letter offers a quantitative evaluation of the impact of mutual coupling on RIS-assisted channel estimation.
Our analysis provides insights into situations where mutual coupling can be disregarded safely.
The analyses and numerical results reveal that within practical scenarios, closer integration of RIS elements or the enlargement of RIS size accentuates the impact of neglecting mutual coupling. 
In addition, even with mutual coupling-aware setups, excessively tight RIS element spacing can lead to substantial degradation in the channel estimation performance.
\end{abstract}


\section{Introduction}

\Ac{ris}, also known as intelligent reflecting surface, empowers intelligent and programmable wireless propagation environments. 
To date, \ac{ris}-assisted communication, sensing, and localization in mmWave/THz band have attracted widespread attention~\cite{Bjornson2022Reconfigurable,Sarieddeen2021Overview,Zheng2023JrCUP,Chen2023Multi}.
Among the \ac{ris}-related studies, channel estimation in a \ac{ris}-assisted wireless system is a particularly essential topic since accurate channel state information is crucial for achieving optimal channel capacity~\cite{Bjornson2020Intelligent}. 
Despite notable advancements in \ac{ris}-assisted channel estimation~\cite{He2021Channel,An2022Low}, the oversight of practical electromagnetic attributes, such as mutual coupling among \ac{ris} unit cells, remains a critical concern.

The mutual coupling effect, which refers to the interaction between neighboring \ac{ris} unit cells, has recently gained attention in \ac{ris}-assisted communications~\cite{Gradoni2021End}.
Generally, mutual coupling among \ac{ris} unit cells can be reasonably ignored when the inter-distance is large enough (e.g., larger than half wavelength~\cite{Qian2021Mutual}).
However, the emergence of techniques such as holographic \ac{mimo} necessitates ultra-dense antenna integration within a limited surface~\cite{An2023Tutorial}, rendering mutual coupling undeniable.
To the best of the authors’ knowledge, mutual coupling-aware channel estimation for \ac{ris}-assisted communications has not yet been studied.
Given this, assessing the justifiable extent of disregarding mutual coupling emerges as a key issue to be addressed.
Using the \ac{mcrb}~\cite{Fortunati2017Performance}, our results show how the system parameters like the inter-distance of \ac{ris} unit cells and the \ac{ris} size affect channel estimation performance with the mutual coupling effect and provide insights into conditions under which mutual coupling can be confidently ignored. The simulation code of this paper is available at  \url{https://github.com/ZPinjun/Communication}.

\section{System Model}

\subsection{The End-to-End \Ac{em} Transfer Model}\label{sec_E2E}
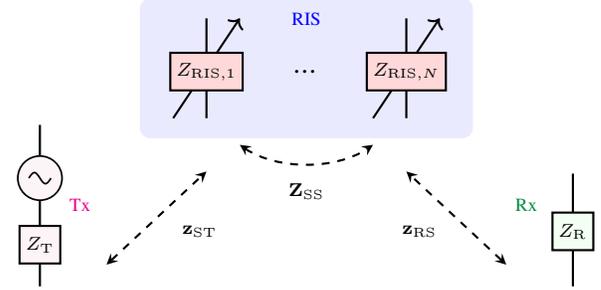
\begin{figure}[t]
    \centering
    \scriptsize
	\begin{circuitikz}[gsn/.style={rectangle, draw=black, fill=green!5, thick, minimum size=5mm},msn/.style={rectangle, draw=black, fill=magenta!5, thick, minimum size=5mm}, rsn/.style={rectangle, draw=black, fill=red!15, thick, minimum size=5mm}]
		\fill[Icol!15, rounded corners] (0.15*\linewidth,0.2*\linewidth) rectangle (0.65*\linewidth,0.41*\linewidth);
		% Tx
		\node at (0.06*\linewidth,0.1*\linewidth){\magenta{Tx}};
		\draw[black, thick] (0,0.22*\linewidth) -- (0,-0.2);
		\draw[line width=0.5, scale=0.7, transform shape] (0,0.3*\linewidth) to [sinusoidal voltage source, fill=magenta!5] (0,0.1*\linewidth);
		\node at (0,0.04*\linewidth)[msn]{$Z_\mathrm{T}$};
		% Rx
		\node at (0.73*\linewidth,0.1*\linewidth){\color[RGB]{0,139,69}Rx};
		\draw[black, thick] (0.8*\linewidth,1.3) -- (0.8*\linewidth,-0.2);
		\node at (0.8*\linewidth,0.06*\linewidth)[gsn]{$Z_\mathrm{R}$};
		% RIS
		\node at (0.4*\linewidth,0.38*\linewidth){\blue{RIS}};
		\draw[black, thick] (0.25*\linewidth,0.23*\linewidth) -- (0.25*\linewidth,0.38*\linewidth);
		\draw[black, thick, ->] (0.20*\linewidth,0.23*\linewidth) -- (0.3*\linewidth,0.38*\linewidth);
		\node at (0.25*\linewidth,0.3*\linewidth)[rsn]{$Z_{\mathrm{RIS},1}$};
		\node at (0.4*\linewidth,0.3*\linewidth){\textbf{\dots}};
		\draw[black, thick] (0.55*\linewidth,0.23*\linewidth) -- (0.55*\linewidth,0.38*\linewidth);
		\draw[black, thick, ->] (0.5*\linewidth,0.23*\linewidth) -- (0.6*\linewidth,0.38*\linewidth);
		\node at (0.55*\linewidth,0.3*\linewidth)[rsn]{$Z_{\mathrm{RIS},N}$};
		% others
		\draw[black, thick, stealth-stealth, dashed] (0.1*\linewidth,0.01*\linewidth) -- (0.25*\linewidth,0.15*\linewidth);
		\node at (0.24*\linewidth,0.06*\linewidth){$\mathbf{z}_{\mathrm{ST}}$};
		\draw[black, thick, stealth-stealth, dashed] (0.7*\linewidth,0.01*\linewidth) -- (0.55*\linewidth,0.15*\linewidth);
		\node at (0.57*\linewidth,0.06*\linewidth){$\mathbf{z}_{\mathrm{RS}}$};
		\draw[black, thick, stealth-stealth, dashed] (0.3*\linewidth,0.19*\linewidth) .. controls (0.35*\linewidth,0.15*\linewidth) and (0.45*\linewidth,0.15*\linewidth) .. (0.5*\linewidth,0.19*\linewidth);
		\node at (0.4*\linewidth,0.12*\linewidth){$\mathbf{Z}_{\mathrm{SS}}$};
	\end{circuitikz}
	\vspace{-1em}
  \caption{
    Schematic illustration of the considered \ac{ris}-assisted communication system with the self and mutual coupling of \ac{ris} elements denoted as $\Zm_\mathrm{SS}$.
  }
  \label{Fig_system}
\end{figure}

We consider an \ac{ris}-assisted \ac{siso} communication system, where the direct path between the transmitter and the receiver is blocked.
The \ac{ris} is assumed to be an array of $N_1\times N_2$ passive yet reconfigurable scattering elements with a uniform inter-distance $d$.
Each of these elements in turn consists of a single unit cell.
To analyze the impact of mutual coupling among scattering elements of the \ac{ris}, we adopt the end-to-end \ac{em}-compliant communication model proposed in~\cite{Gradoni2021End}, which is given by
 \begin{equation}
 	h_{\mathrm{E2E}} = \zv_{\Rt\St}^\TT ( \Zm_{\St\St} + \Zm_{\Rt\It\St} )^{-1}\zv_{\mathrm{ST}}.
 \end{equation}
 Here, $\zv_{\mathrm{ST}}\in\mathbb{C}^{N\times 1}$ denotes the mutual impedances between the transmitter and \ac{ris}, $\zv_{\Rt\St}\in\mathbb{C}^{N\times 1}$ denotes the mutual impedances between the \ac{ris} and receiver,~$\Zm_{\St\St}\in\mathbb{C}^{N\times N}$ is a full matrix containing the self and mutual impedances among scattering elements of the \ac{ris}, and $\Zm_{\Rt\It\St}=\mathrm{diag}\{Z_{\mathrm{RIS},1},Z_{\mathrm{RIS},2},\dots,Z_{\mathrm{RIS},N}\}\in\mathbb{C}^{N\times N}$ is a diagonal matrix containing the tunable loads~$Z_{\mathrm{RIS},n}$ of the circuits enabling the reconfigurability of the surface, $n=1,2,\dots,N_1\times N_2$.
 A schematic diagram of the system is provided in Fig.~\ref{Fig_system}.
In this model, the wireless channels correspond to the impedance vectors~$\zv_{\Rt\St}$ and~$\zv_{\mathrm{ST}}$.

Note that based on the microwave network theory~\cite{Shen2022Modeling}, the commonly employed channel coefficients are essentially scattering parameters, which are equivalent characterizations to the impedance vectors discussed in this letter. The conversion between these two representations is elucidated in~\cite[Eq.~(6)]{Li2023Beyond} and~\cite[Eq.~(35)--(39)]{Shen2022Modeling}.
We can further decompose~$
	\Zm_{\St\St} = \Zm_{\St\St}^{\mathrm{self}} + \Zm_{\St\St}^{\mathrm{mutual}},
$
where $\Zm_{\St\St}^{\mathrm{self}}\in\mathbb{C}^{N\times N}$ is a diagonal matrix representing the self impedances of the \ac{ris} elements, and $\Zm_{\St\St}^{\mathrm{mutual}}$ is a matrix containing only the off-diagonal entries which stand for the mutual impedances among each pair of \ac{ris} elements.
Hence, the matrix~$\Zm_{\St\St}^{\mathrm{mutual}}$ characterizes the mutual coupling among \ac{ris} elements.

\begin{figure}[t]
    \centering
	\begin{tikzpicture}
	\begin{axis}[%
	width=2.5in, height=1in, at={(0,0)}, scale only axis, 
	xmin=-0.1, xmax=2.5,
	xlabel style={font=\color{white!15!black},font=\footnotesize},
  	xticklabel style = {font=\color{white!15!black},font=\footnotesize},
	xlabel={Distance between two radiators $(\lambda)$},
	ymode=log, ymin=0.01, ymax=1000, yminorticks=true,
	ylabel style={font=\color{white!15!black},font=\footnotesize},
  	yticklabel style = {font=\color{white!15!black},font=\footnotesize},
  	ytick = {0.01,0.1,1,10,100,1000},
	ylabel={$|Z_{qp}|\ (\Omega)$},
	axis background/.style={fill=white},
	xmajorgrids, ymajorgrids,
	]
	\addplot [color=blue, line width=0.8pt, forget plot]
  	table[row sep=crcr]{%
	0.002	1510.22956938793\\
	0.005	699.652223168663\\
	0.01	286.51210061345\\
	0.02	79.1291177214507\\
	0.05	7.74798094422523\\
	0.1	0.977518169863381\\
	0.111111111111111	0.712980086772819\\
	0.125	0.509137710675962\\
	0.142857142857143	0.361242223647806\\
	0.166666666666667	0.262388739773888\\
	0.2	0.200867432578507\\
	0.25	0.160237412859006\\
	0.333333333333333	0.125383938560281\\
	0.5	0.087849504032253\\
	1	0.0455164537678996\\
	2.5	0.018400205890376\\
	};
	\end{axis}
	\end{tikzpicture}
  	\caption{
    An example of $|Z_{qp}|$ versus the inter-distance of two radiators with $h_p=h_q=\lambda/64$ and $r=\lambda/500$ at frequency of \unit[28]{GHz}.
  	}
  	\label{Fig_zqp}
\end{figure}
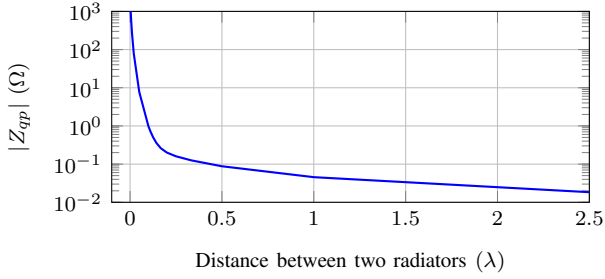

According to~\cite{Gradoni2021End}, the mutual impedances $\{\zv_{\mathrm{ST}},\zv_{\mathrm{RS}},\Zm_{\St\St}\}$ depend only on the physical configuration of the radiators.
When it is assumed that all radiators, including antennas and scatterers, are cylindrical thin wires of perfectly conducting material, the mutual impedances can be calculated explicitly~\cite{Di2023Modeling}.
For two cylindrical thin radiators (each with radius~$r$) located at $\pv_p=[x_p,y_p,z_p]^\TT$ and $\pv_q=[x_q,y_q,z_q]^\TT$, the mutual impedance can be evaluated numerically by~\eqref{eq_MI} given below~\cite{Gradoni2021End,Di2023Modeling}.
\begin{figure*}[b]
\hrulefill
{\small
	\begin{multline}\label{eq_MI}
		Z_{qp} = \frac{j\eta_0}{4\pi k_0}\int_{-h_q}^{h_q}\int_{-h_p}^{h_p}\frac{e^{-jk_0R(\xi,z)} \sin(k_0(h_p-|\xi|))\sin(k_0(h_q-|z|))}{R(\xi,z)\sin(k_0h_p)\sin(k_0h_q)} \bigg( k_0^2 - \frac{jk_0}{R(\xi,z)} - \frac{k_0^2(z-\xi+\rho_2)^2+1}{R^2(\xi,z)} \\
		+ \frac{3jk_0(z-\xi+\rho_2)^2}{R^3(\xi,z)} + \frac{3(z-\xi+\rho_2)^2}{R^4(\xi,z)} \bigg) \mathrm{d} \xi \mathrm{d} z,
	\end{multline}
}
\end{figure*}
Here, $\eta_0=\sqrt{{\mu_0}/{\epsilon_0}}$ and $k_0={2\pi}/{\lambda}$ are the intrinsic impedance of free space and the wavenumber, where $\mu_0$, $\epsilon_0$, and $\lambda$ denote the magnetic permeability, the electric permittivity, and the signal wavelength, respectively.
The integral limit variables~$h_p$ and $h_q$ represent the half-length of the two radiators and the function $R(\xi,z)$ accounting for the distance between $\pv_p$ and $\pv_q$ is given by
\begin{align}
	&R(\xi,z) = \sqrt{\rho_1^2 + (z-\xi+\rho_2)^2 },\\
	&\begin{cases}
		\rho_1 =\sqrt{(x_p-x_q)^2 + (y_p-y_q)^2},\ \rho_2 =  z_p-z_q, &\text{if } p\!\neq\! q,\\
		\rho_1 =r,\ \rho_2 = 0,  &\text{if } p\! =\! q,
	\end{cases}\notag
\end{align}
Here $p = q$ corresponds to the self-impedance case.
Note that a shorter distance between $\pv_p$ and $\pv_q$ generates a lower value of $R(\xi,z)$, thus resulting in a higher mutual impedance according to~\eqref{eq_MI}.
Fig.~\ref{Fig_zqp} plots a typical example of mutual coupling $|Z_{qp}|$ versus the inter-distance of two radiators.

\subsection{Signal Model}

In this work, we consider a scenario of an uplink transmission from an unknown user to a fixed base station through a fixed \ac{ris}.
Suppose the transmitter sends a deterministic pilot signal $\sqrt{P_\mathrm{T}}\cdot 1$, with power $P_\mathrm{T}$, to the receiver via \ac{ris}.
The received signal can be expressed as
\begin{equation}\label{eq_y}
	y = \sqrt{P_\mathrm{T}}\zv_{\Rt\St}^\TT ( \Zm_{\St\St} + \Zm_{\Rt\It\St} )^{-1}\zv_{\mathrm{ST}} + \omega,
\end{equation}
where $\omega\sim\mathcal{CN}(0,\sigma^2)$ models the thermal noise at the receiver. 
Since the locations of the receiver and \ac{ris} are known, the RIS-receiver channel~$\zv_{\Rt\St}$ is assumed to be known.
Hence, the channel estimation problem in this paper focuses on the estimation of the unknown channel~$\zv_{\mathrm{ST}}$.

The signal model~\eqref{eq_y} indicates that the observable $y$ is a projection of~$\zv_{\mathrm{ST}}$  onto the direction of the vector~$\sqrt{P_\mathrm{T}}\zv_{\Rt\St}^\TT ( \Zm_{\St\St} + \Zm_{\Rt\It\St} )^{-1}$ which depends on the \ac{ris} configuration $\Zm_{\Rt\It\St}$ in addition to the mutual coupling $\Zm_{\St\St}$.
To estimate the unknown channel~$\zv_{\mathrm{ST}}$, multiple transmissions are required and the \ac{ris} needs to switch between different configurations so that the projections of~$\zv_{\mathrm{ST}}$ in different directions are observable~\cite{Bjornson2022Maximum}.
Suppose $G$ transmissions are implemented at different time instants, during which the channel~$\zv_{\mathrm{ST}}$ stays constant and the \ac{ris} configuration is changed over these instances as $\Zm_{\Rt\It\St,1}$, $\Zm_{\Rt\It\St,2}$, \dots, $\Zm_{\Rt\It\St,G}$.
Then the total received signal can be concatenated as $\yv=[y_1,\dots,y_G]^\TT\in\mathbb{C}^{G\times 1}$, where $y_g = \sqrt{P_\mathrm{T}}\zv_{\Rt\St}^\TT ( \Zm_{\St\St} + \Zm_{\Rt\It\St,g} )^{-1}\zv_{\mathrm{ST}} + \omega$, $g=1,\dots,G$.

\subsection{Problem Formulation and Analysis Objective}

\subsubsection{The Channel Estimation Problem}
The goal of this letter is to analyze the impact of mutual coupling on the channel estimation of RIS-aided communication. Hence, we first clarify the considered mismatched channel estimation problem. To begin with, we define
\begin{equation}
	\Bm(\Zm_{\St\St}^{\mathrm{self}},\Zm_{\St\St}^{\mathrm{mutual}}) \stackrel{\triangle}{=} \begin{bmatrix}
		\zv_{\Rt\St}^\TT ( \Zm_{\St\St}^{\mathrm{self}} + \Zm_{\St\St}^{\mathrm{mutual}} + \Zm_{\Rt\It\St,1} )^{-1} \\
		\vdots\\
		\zv_{\Rt\St}^\TT ( \Zm_{\St\St}^{\mathrm{self}} + \Zm_{\St\St}^{\mathrm{mutual}} + \Zm_{\Rt\It\St,G} )^{-1}
	\end{bmatrix}.\notag
\end{equation}
Then, the observations, i.e., the received signal with RIS mutual coupling, can be written as
\begin{equation}\label{eq_yvobs}
	\yv\! =\! \sqrt{P_\mathrm{T}}\Bm(\Zm_{\St\St}^{\mathrm{self}},\Zm_{\St\St}^{\mathrm{mutual}}) \zv_{\mathrm{ST}} + \omegav,
\end{equation}
where $\omegav\sim\mathcal{CN}(\mathbf{0},\sigma^2\mathbf{I}_G)$.
Next, we consider a channel estimation problem using the mutual coupling-unaware model given by~\cite{Qian2021Mutual}
\begin{equation}\label{eq_yvest}
	\yv = \sqrt{P_\mathrm{T}}\Bm(\Zm_{\St\St}^{\mathrm{self}},\mathbf{0})  \zv_{\mathrm{ST}} + \omegav.
\end{equation}
\textit{The considered channel estimation refers to estimating~$\zv_{\mathrm{ST}}$ based on the received signal~$\yv$, where~$\yv$ is given by the true model~\eqref{eq_yvobs} while the estimation is based on model~\eqref{eq_yvest} that does not account for mutual coupling.}

\subsubsection{The Analysis Objective}
Since we neglect the mutual coupling, the estimation model~\eqref{eq_yvest} is mismatched with the true model~\eqref{eq_yvobs}. Thus, our analyses aim to offer insights into how this mismatch affects the estimation performance.

We further express these quantities in terms of the real and imaginary components as
\begin{align}\label{eq_Dbar}
	&\Dm(\Zm_{\St\St}^{\mathrm{self}},\Zm_{\St\St}^{\mathrm{mutual}}) = \begin{bmatrix}
	\mathrm{Re}(\Bm) & -\mathrm{Im}(\Bm)\\
	\mathrm{Im}(\Bm) & \mathrm{Re}(\Bm)
	\end{bmatrix}\in\mathbb{R}^{2G\times 2N},\\
	&\rv\! =\! \begin{bmatrix}
	\mathrm{Re}(\yv^\TT),\ 
	\mathrm{Im}(\yv^\TT)
	\end{bmatrix}^\TT,\quad
	\xv\! =\! \begin{bmatrix}
	\mathrm{Re}(\zv_{\mathrm{ST}}^\TT),
	\mathrm{Im}(\zv_{\mathrm{ST}}^\TT)
	\end{bmatrix}^\TT,
\end{align}
where $\mathrm{Re}(\cdot)$ and $\mathrm{Im}(\cdot)$ respectively return the real and imaginary parts of a complex vector or matrix.
Thus, the observation model~\eqref{eq_yvobs} and the estimation model~\eqref{eq_yvest} can be rewritten as 
\begin{align}
\textit{Obs. model:\ }&	\rv\! =\! \sqrt{P_\mathrm{T}} \Dm(\Zm_{\St\St}^{\mathrm{self}},\Zm_{\St\St}^{\mathrm{mutual}})\xv + \boldsymbol{\varpi} \in\mathbb{R}^{2G}, \\
\textit{Est. model:\ }&	\rv\! =\! \sqrt{P_\mathrm{T}} \Dm(\Zm_{\St\St}^{\mathrm{self}},\mathbf{0})\xv + \boldsymbol{\varpi} \in\mathbb{R}^{2G}, 
\end{align}
where $\boldsymbol{\varpi}\sim\mathcal{N}(\mathbf{0},\frac{\sigma^2}{2}\mathbf{I}_{2G})$.
Hence, the channel estimation problem can be equally stated as recovering $\xv$ given $\rv$.

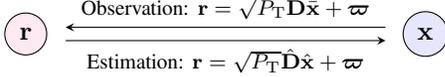
\begin{figure}[t]
    \centering
    \vspace{-0.5em}
	\begin{tikzpicture}[bsn/.style={circle, draw=black, fill=blue!10, thin, minimum size=5mm},msn/.style={circle, draw=black, fill=magenta!10, thin, minimum size=5mm}]
		\node at (0.2*\linewidth,0.1*\linewidth)[msn]{$\rv$};
		\node at (0.8*\linewidth,0.1*\linewidth)[bsn]{$\xv$};
		\draw[black, thin, stealth-] (0.26*\linewidth,0.11*\linewidth) -- (0.74*\linewidth,0.11*\linewidth);
		\draw[black, thin, -stealth] (0.26*\linewidth,0.09*\linewidth) -- (0.74*\linewidth,0.09*\linewidth);
		\node at (0.5*\linewidth,0.14*\linewidth){\footnotesize Observation: $\rv=\sqrt{P_\mathrm{T}}\bar{\Dm}\bar{\xv} + \boldsymbol{\varpi}$};
		\node at (0.5*\linewidth,0.06*\linewidth){\footnotesize Estimation: $\rv=\sqrt{P_\mathrm{T}}\hat{\Dm}\hat{\xv} + \boldsymbol{\varpi}$};
	\end{tikzpicture}
	\vspace{-0.5em}
  \caption{
    Illustration of the mutual coupling-introduced model mismatch in the considered \ac{ris}-assisted channel estimation problem.
  }
  \label{Fig_mismatch}
\end{figure}

For the sake of better illustration, we denote  
\begin{align}
	\bar{\Dm}=\Dm(\Zm_{\St\St}^{\mathrm{self}},\Zm_{\St\St}^{\mathrm{mutual}}),\quad
	\hat{\Dm}=\Dm(\Zm_{\St\St}^{\mathrm{self}},\mathbf{0}). \label{eq_Dtilde}
\end{align}
In addition, we use $\bar{\xv}$ and $\hat{\xv}$ to represent the true unknown channel and the estimated channel through the mismatched model that ignores the \ac{ris} mutual coupling, respectively.
Then the involved model mismatch can be illustrated as Fig.~\ref{Fig_mismatch}.
\textit{We are interested in the behavior of the estimation \ac{rmse} $\sqrt{\mathbb{E}\|\hat{\xv}-\bar{\xv}\|^2}$ under this model mismatch, which can be analyzed using the tool of \ac{mcrb}.}

\section{\ac{mcrb} Analysis}\label{sec_MCRB}

The \ac{mcrb} is a generalization of the well-known \ac{crlb} to the cases with a model mismatch, which gives a lower bound for the estimation \ac{rmse}~\cite{Fortunati2017Performance,Zheng2023Misspecified}. Based on Fig.~\ref{Fig_mismatch}, we have the true likelihood function $f_\mathrm{T}$ and the mismatched likelihood function $f_\mathrm{M}$ as
\begin{align}
	f_\mathrm{T}(\rv;\xv) &= \frac{1}{(\pi\sigma^2)^G}e^{-\frac{\|\rv -\sqrt{P_\mathrm{T}}\bar{\Dm}{\xv}\|^2}{\sigma^2}},\\  
	f_\mathrm{M}(\rv;\xv) &= \frac{1}{(\pi\sigma^2)^G}e^{-\frac{\|\rv -\sqrt{P_\mathrm{T}}\hat{\Dm}{\xv}\|^2}{\sigma^2}}.\label{eq_fM}
\end{align}
Based on model $f_\mathrm{M}$, the lower bound matrix of the estimation \ac{mse} can be obtained as~\cite{Fortunati2017Performance}
\begin{equation}\label{eq_LBM}
	\mathrm{LBM}(\hat{\xv},\bar{\xv}) = \underbrace{\Am_{\xv_0}^{-1} \Bm_{\xv_0} \Am_{\xv_0}^{-1}}_{\mathrm{MCRB}(\xv_0)} + \underbrace{(\bar{\xv}-\xv_0)(\bar{\xv}-\xv_0)^\TT}_{\mathrm{Bias}(\xv_0)},
\end{equation}
thus the \ac{rmse} of the channel estimation under model mismatch is lower bounded as
\begin{equation}\label{eq_LB}
	\sqrt{\mathbb{E}\|\hat{\xv}-\bar{\xv}\|^2} \geq \sqrt{\mathrm{Tr}(\mathrm{LBM}(\hat{\xv},\bar{\xv}))} \triangleq \mathrm{LB},
\end{equation}
where $\mathrm{Tr}(\cdot)$ returns the trace of a matrix. 
Here, $\xv_0$ is the pseudo-true parameter vector that minimizes the \ac{kld} between $f_\mathrm{T}$ and $f_\mathrm{M}$, i.e., 
\begin{equation}
	\xv_0 = \arg\min_{\xv}\ \mathrm{KLD}\left(f_\mathrm{T}(\rv;\bar{\xv})||f_\mathrm{M}(\rv;\xv)\right),
\end{equation}
and matrices $\Am_{\xv_0}$ and $\Bm_{\xv_0}$ are defined as 
\begin{align}
	\Am_{\xv_0}\!&=\!\mathbb{E}_{f_\mathrm{T}}\bigg\{\frac{\partial^2}{\partial\xv^2}\ln f_\mathrm{M}(\rv;\xv)\bigg|_{\xv=\xv_0}\bigg\},
	\label{eq_Ax0}\\
	\Bm_{\xv_0}\!&=\!\mathbb{E}_{f_\mathrm{T}}\bigg\{\frac{\partial\ln f_\mathrm{M}(\rv;\xv)}{\partial\xv} \bigg(\frac{\partial\ln f_\mathrm{M}(\rv;\xv)}{\partial\xv}\bigg)^\TT\bigg|_{\xv=\xv_0} \bigg\}.
	\label{eq_Bx0}
\end{align}

\subsection{\ac{mcrb} Derivation}

Now we focus on evaluating $\xv_0$, $\Am_{\xv_0}$, and $\Bm_{\xv_0}$ that appear in~\eqref{eq_LBM}.
According to the definition of \ac{kld}~\cite{Sayed2022Inference}, we can write 
\begin{align}
	&\mathrm{KLD}\left(f_\mathrm{T}(\rv;\bar{\xv})||f_\mathrm{M}(\rv;\xv)\right)
	= \mathbb{E}_{f_\mathrm{T}}\left\{\ln f_\mathrm{T}(\rv;\bar{\xv}) - \ln f_\mathrm{M}(\rv;\xv)\right\},\notag\\
	&=\!-\!\frac{1}{\sigma^2}\mathbb{E}_{f_\mathrm{T}}\!\left\{\|\rv-\sqrt{P_\mathrm{T}}\bar{\Dm}\bar{\xv}\|^2\right\}\!  +\! \frac{1}{\sigma^2}\mathbb{E}_{f_\mathrm{T}}\!\left\{\|\rv-\sqrt{P_\mathrm{T}}\hat{\Dm}{\xv}\|^2\right\},\notag\\
	&= \frac{P_\mathrm{T}}{\sigma^2}\|\bar{\Dm}\bar{\xv}-\hat{\Dm}\xv\|^2.
\end{align}
It thus follows that
\begin{align}\label{eq_x0de}
	\xv_0 = \arg\min_{\xv}\ \|\bar{\Dm}\bar{\xv}-\hat{\Dm}\xv\|^2
	= \big(\hat{\Dm}^\TT\hat{\Dm}\big)^{-1}\hat{\Dm}^\TT\bar{\Dm}\bar{\xv}.
\end{align}
Substituting $f_\mathrm{M}$ from~\eqref{eq_fM} in~\eqref{eq_Ax0} and~\eqref{eq_Bx0} yields
\begin{align}\label{eq_Ax0Bx0de}
	\Am_{\xv_0} = -\Bm_{\xv_0}= -\frac{2P_\mathrm{T}}{\sigma^2}\hat{\Dm}^{\TT}\hat{\Dm}.
\end{align}
By substituting~\eqref{eq_x0de} and~\eqref{eq_Ax0Bx0de} into~\eqref{eq_LBM}, we finally obtain
\begin{equation}\label{eq_LBMderived}
	\mathrm{LBM}(\hat{\xv},\bar{\xv}) = \underbrace{\frac{1}{2\gamma}\big(\hat{\Dm}^{\TT}\hat{\Dm}\big)^{-1}}_{\mathrm{MCRB}(\xv_0)} + \underbrace{\big(\bar{\xv}-\Phim\bar{\xv}\big)\big(\bar{\xv}-\Phim\bar{\xv}\big)^\TT}_{\mathrm{Bias}(\xv_0)},
\end{equation}
where $\Phim=\big(\hat{\Dm}^\TT\hat{\Dm}\big)^{-1}\hat{\Dm}^\TT\bar{\Dm}$ and $\gamma\triangleq P_\mathrm{T}/\sigma^2$ is the \ac{snr}.

\subsection{Results Analysis}

\begin{figure}[t]
    \centering
    \definecolor{ForestGreen}{rgb}{0.1333    0.5451    0.1333}
	\begin{tikzpicture}
	\begin{axis}[%
	width=0.6*\linewidth,
	height=0.5*\linewidth,
	at={(0,0)},
	scale only axis,
	xmin=-4.5,
	xmax=5.5,
	ymin=-4,
	ymax=6,
	xtick={\empty},
	ytick={\empty},
	axis background/.style={fill=white},
	legend style={at={(1,0)}, anchor=south east, legend cell align=left, align=left, draw=white!15!black, font=\tiny}
	]
	\addplot[only marks, mark=+, line width=1pt, mark size=3pt, draw=black] table[row sep=crcr]{%
	x	y\\
	-1.406939998315023	-2.63585356339807\\
	-3.54817257531216	-3.33291731004782\\
	2.9655083892214	1.51271718754857\\
	-3.0688314835517	-2.46137038403269\\
	4.89560115750507	4.66896540698647\\
	2.01567697076346	5.18133121466441\\
	-3.28992221532228	4.14996741739059\\
	4.80392748258537	-0.0817247704107086\\
	-4.02821011003632	2.47581234347691\\
	-2.45268624093629	7.21225404286588\\
	}; \addlegendentry{Estimates $\hat{\xv}$ under low SNR}
	\addplot[only marks, mark=x, line width=1pt, mark size=3pt, draw=ForestGreen] table[row sep=crcr]{%
	x	y\\
	0.0105986622585124	-0.421988086368854\\
	-0.0355780173668363	-0.159879025201191\\
	0.354496512391082	0.113395018579846\\
	-0.501760722616177	-0.000288579772512069\\
	-0.0913178370718468	0.124784933250949\\
	0.486075507387476	0.0252881662069351\\
	-0.0942908171797427	0.136173558837508\\
	-0.112065698229665	-0.257483182772147\\
	-0.245293474752728	0.0436169705049455\\
	0.158590072719643	-0.31331341473916\\
	}; \addlegendentry{Estimates $\hat{\xv}$ under high SNR}
	\addplot[only marks, mark=*, mark options={solid,blue}, mark size=2.5pt, draw=blue] table[row sep=crcr]{%
	x	y\\
	0	0\\
	}; \addlegendentry{Pseudo-true parameter ${\xv}_0$}
	\addplot[only marks, mark=*, mark options={solid, red}, mark size=2.5000pt, draw=red] table[row sep=crcr]{%
	x	y\\
	1	1\\
	}; \addlegendentry{True unknown parameter $\bar{\xv}$}
	\draw[blue, thick, stealth-stealth] (axis cs: -3.12992221532228, 3.90996741739059) -- (axis cs: -0.2, 0.2);
	\draw[red, thick, stealth-stealth] (axis cs: 0.15, 0.15) -- (axis cs: 0.85, 0.85);
	\draw[black, thick, stealth-stealth] (axis cs: -3.06992221532228, 3.96996741739059) -- (axis cs: 0.8, 1.2);
	\node[right, align=left] at (axis cs:-4.5,1) {\tiny{\blue{$\sqrt{\mathbb{E}\|\hat{\xv}-\xv_0\|^2}$}}\\\tiny{\blue{$\geq\sqrt{\mathrm{Tr}(\mathrm{MCRB})}$}}};
	\node[left, align=left] at (axis cs:3,0.3) {\tiny{\red{$\sqrt{\mathrm{Tr}(\mathrm{Bias})}$}}};
	\node[left, align=left] at (axis cs:3.5,3.2) {\colorbox{yellow!50}{\tiny{$\sqrt{\mathbb{E}\|\hat{\xv}-\bar{\xv}\|^2}\geq\mathrm{LB}$}}};
\end{axis}
\end{tikzpicture}
  \caption{
    Illustration of the relationship between $\mathrm{MCRB}$, $\mathrm{Bias}$, and $\mathrm{LB}$.
  }
  \label{Fig_MCRBanalysis}
\end{figure}
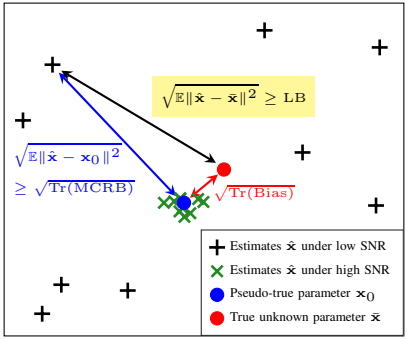

Fig.~\ref{Fig_MCRBanalysis} provides an intuitive illustration of~\eqref{eq_LBM} and~\eqref{eq_LB}.
Under a certain model mismatch, the estimates (e.g., using a maximum likelihood estimator) converge to the pseudo-true parameters as the \ac{snr} increases~\cite{Fortunati2017Performance}.
As shown in Fig.~\ref{Fig_MCRBanalysis}, the $\mathrm{MCRB}$ term in~\eqref{eq_LBM} lower bounds the \ac{mse} between the estimates $\hat{\xv}$ and the pseudo-true parameter $\xv_0$, while the $\mathrm{Bias}$ term accounts for the distance between pesudo-true parameter $\xv_0$ and the true unknown parameter $\bar{\xv}$.
Based on the results in~\eqref{eq_LBMderived}, the following insights are further gained. 

\begin{remark}
	The $\mathrm{MCRB}$ term in~\eqref{eq_LBMderived} decreases with increasing \ac{snr} and finally converges to zero.
	The value of $\mathrm{MCRB}$ at a certain \ac{snr} depends only on the mismatched model $\hat{\Dm}$ itself, and is independent of the true model $\bar{\Dm}$ and the divergence between the true and mismatched models.
\end{remark}
\begin{remark}
	The $\mathrm{Bias}$ term in~\eqref{eq_LBMderived} is independent of \ac{snr}. 
	Instead, it depends only on the mismatch between $\bar{\Dm}$ and $\hat{\Dm}$.
\end{remark}
\begin{remark}\label{remark_sat}
	In the low \ac{snr} region where $\mathrm{MCRB}$ is dominant, $\mathrm{LB}$ can be approximated by~$\sqrt{\mathrm{Tr}(\mathrm{MCRB})}$; While in the high \ac{snr} region where $\mathrm{MCRB}$ converges to zero, $\mathrm{LB}$ will saturate to $\sqrt{\mathrm{Tr}(\mathrm{Bias})}$, as demonstrated in Fig.~\ref{Fig_MCRBanalysis}.
\end{remark}
\begin{remark}\label{remark_4}
	When we estimate using the true model, i.e.,~$\hat{\Dm}=\bar{\Dm}$, the mismatched $\mathrm{LB}$ is reduced to the classical \ac{crlb}, which is obtained as~$\text{CRLB}=\sqrt{\mathrm{Tr}(\mathrm{MCRB})}=\sqrt{\frac{1}{2\gamma}\mathrm{Tr}\{\bar{\Dm}^\TT\bar{\Dm}\}}$ from~\eqref{eq_LBMderived}.
	Remarkably, according to \textit{Remark~1}, we can infer that the estimation \ac{mse} using the correct model does not necessarily outperform the mismatched model in the $\mathrm{MCRB}$-dominant regions, since the $\mathrm{MCRB}$ term depends only on the used model itself.
\end{remark}

\section{Numerical Results}
This section reports numerical results. 
The simulations setups are as follows: $h_p=h_q=\lambda/64$, $r=\lambda/500$, the signal frequency $f=\unit[28]{GHz}$, the position of the transmitter is $(\unit[5]{m},\unit[-5]{m},\unit[3]{m})$, the position of the receiver is $(\unit[5]{m},\unit[5]{m},\unit[1]{m})$, and the \ac{ris} is centered at $(\unit[0]{m},\unit[0]{m},\unit[0]{m})$, which are the same as~\cite{Gradoni2021End}.
In addition, the number of transmissions $G=256$.
For the $g$th transmission, we set the tunable load of the $n$th \ac{ris} element as $Z_{\mathrm{RIS},n,g}=R_{\mathrm{RIS},n,g}+j\omega L_{\mathrm{RIS},n,g}$, where $\omega=2\pi f$, $R_{\mathrm{RIS},n,g}$ is randomly generated from $[\unit[0.1]{\Omega},\ \unit[10.1]{\Omega}]$, and $L_{\mathrm{RIS},n,g}$ is randomly generated from $[\unit[0.1]{nH},\ \unit[10.1]{nH}]$.
By default, the \ac{ris} size is $4\times 4$, the noise PSD is~\unit[-173.855]{dBm/Hz}, and the noise figure is~\unit[10]{dB}.

\subsection{Impact of Ignoring Mutual Coupling}\label{sec_sim1}

This subsection evaluates the impact of ignoring mutual coupling on \ac{ris}-assisted channel estimation.
We first study the behavior of the derived bounds at different \ac{snr} levels. 
To verify our bound derivation, we test the \acp{rmse} of the \ac{ml} estimator $\hat{\xv}=\frac{1}{\sqrt{P_\mathrm{T}}}(\hat{\Dm}^\TT\hat{\Dm})^{-1}\hat{\Dm}^\TT\rv$ and compare them with the derived bounds.
The observation~$\rv$ is generated based on the true model~$\bar{\Dm}$.
As clarified in Section~\ref{sec_E2E}, mutual coupling relies on \ac{ris} element spacing $d$, which affects the mutual impedance matrix $\Zm_{\mathrm{SS}}^{\mathrm{mutual}}$ and in turn varies the true model $\bar{\Dm}$ according to~\eqref{eq_Dbar}.
For the sake of clarity, we use $\bar{\Dm}(d)=\Dm\big(\Zm_{\St\St}^{\mathrm{self}},\Zm_{\St\St}^{\mathrm{mutual}}(d)\big)$ to indicate that $\bar{\Dm}$ is a function of $d$.
In contrast, $\hat{\Dm}$ is unrelated to $d$ since it ignores the mutual coupling effect as shown in~\eqref{eq_Dtilde}.

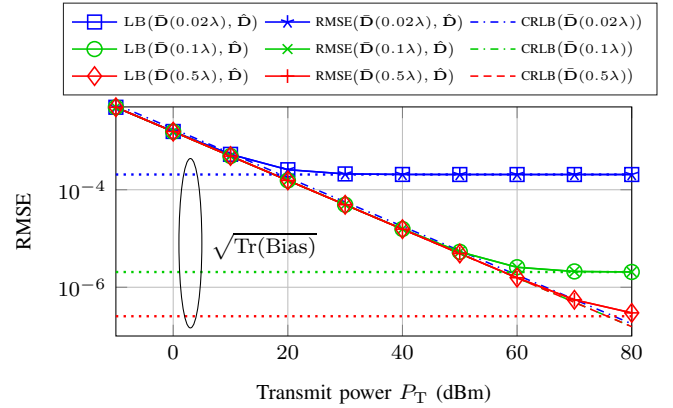
\begin{figure}[t]
    \centering
    \definecolor{ForestGreen}{RGB}{0    204    0}
	\begin{tikzpicture} 
    \begin{axis}[%
	width=2.7in, height=1.2in, at={(0,0)}, scale only axis, 
    xmin=-10, xmax=80, xlabel={Transmit power $P_\mathrm{T}$ (dBm)},
    xlabel style={font=\color{white!15!black},font=\footnotesize},
  	xticklabel style = {font=\color{white!15!black},font=\footnotesize},
    ymin=1e-07,ymax=0.005,ymode=log,yminorticks=true, ylabel={RMSE},
    ylabel style={font=\color{white!15!black},font=\footnotesize},
  	yticklabel style = {font=\color{white!15!black},font=\footnotesize},
    axis background/.style={fill=white},
    xmajorgrids, ymajorgrids,
    legend style={at={(1.05,1.05)}, anchor=south east, legend columns=3, legend cell align=left, align=left, font=\tiny, draw=white!15!black}
    ]
        \addplot [color=blue, line width=0.6pt, mark=square, mark options={solid, blue}, mark size=2.6pt]
      table[row sep=crcr]{%
	-10	0.00492041308218351\\
	0	0.00156811921517313\\
	10	0.000532776848560866\\
	20	0.000257558731358836\\
	30	0.000211151682619025\\
	40	0.00020593660764957\\
	50	0.00020540781801078\\
	60	0.00020535486415696\\
	70	0.000205349568020536\\
	80	0.000205349038399381\\
    };
    \addlegendentry{$\mathrm{LB}\big(\bar{\Dm}(0.02\lambda),\hat{\Dm}\big)$}
    
    \addplot [color=blue, line width=0.6pt, mark=star, mark options={solid, blue}, mark size=2.6pt]
      table[row sep=crcr]{%
	-10	0.00491862778435564\\
	0	0.00157744484214069\\
	10	0.000532093305597625\\
	20	0.000257947025477356\\
	30	0.000210943219496276\\
	40	0.00020600506700055\\
	50	0.000205375524911666\\
	60	0.000205344428795799\\
	70	0.000205351834909312\\
	80	0.000205349287298051\\
    };
    \addlegendentry{RMSE$\big(\bar{\Dm}(0.02\lambda),\hat{\Dm}\big)$}
    
    \addplot [color=blue, line width=0.6pt, dashdotted]
      table[row sep=crcr]{%
 	-10	0.00562014915228002\\
	0	0.00177724721110694\\
	10	0.000562014915228002\\
	20	0.000177724721110694\\
	30	5.62014915228002e-05\\
	40	1.77724721110694e-05\\
	50	5.62014915228002e-06\\
	60	1.77724721110694e-06\\
	70	5.62014915228002e-07\\
	80	1.77724721110694e-07\\
    };
    \addlegendentry{CRLB$\big(\bar{\Dm}(0.02\lambda)\big)$}
    
    \addplot [color=ForestGreen, line width=0.6pt, mark=o, mark options={solid, ForestGreen}, mark size=2.8pt]
      table[row sep=crcr]{%
	-10	0.00491612661661784\\
	0	0.00155461695741779\\
	10	0.000491616905382831\\
	20	0.000155475115034262\\
	30	4.92041092652794e-05\\
	40	1.56811245118711e-05\\
	50	5.32756939828381e-06\\
	60	2.57517546338161e-06\\
	70	2.11101443968769e-06\\
	80	2.05885096456899e-06\\
    };
    \addlegendentry{$\mathrm{LB}\big(\bar{\Dm}(0.1\lambda),\hat{\Dm}\big)$}
    
   	\addplot [color=ForestGreen, line width=0.6pt, mark=x, mark options={solid, ForestGreen}, mark size=2.6pt]
      table[row sep=crcr]{%
	-10	0.00491493177442237\\
	0	0.00156373594846416\\
	10	0.000489535996979962\\
	20	0.00015558757678832\\
	30	4.94274445327616e-05\\
	40	1.56853543010885e-05\\
	50	5.31980693370016e-06\\
	60	2.56366579699844e-06\\
	70	2.10262105466063e-06\\
	80	2.05875836614379e-06\\
    };
    \addlegendentry{RMSE$\big(\bar{\Dm}(0.1\lambda),\hat{\Dm}\big)$}
    
   	\addplot [color=ForestGreen, line width=0.6pt, dashdotted]
      table[row sep=crcr]{%
	-10	0.0049251814187934\\
	0	0.00155747911729268\\
	10	0.00049251814187934\\
	20	0.000155747911729268\\
	30	4.9251814187934e-05\\
	40	1.55747911729268e-05\\
	50	4.9251814187934e-06\\
	60	1.55747911729268e-06\\
	70	4.9251814187934e-07\\
	80	1.55747911729268e-07\\
    };
    \addlegendentry{CRLB$\big(\bar{\Dm}(0.1\lambda)\big)$}
    
    \addplot [color=red, line width=0.6pt, mark=diamond, mark options={solid, red}, mark size=3.5pt]
      table[row sep=crcr]{%
	-10	0.004916126194538\\
	0	0.00155461562268465\\
	10	0.000491612684602856\\
	20	0.000155461768287746\\
	30	4.91619199464929e-05\\
	40	1.55482368837314e-05\\
	50	4.92270250262509e-06\\
	60	1.57528819681474e-06\\
	70	5.53521634980144e-07\\
	80	2.98113284051242e-07\\
    };
    \addlegendentry{$\mathrm{LB}\big(\bar{\Dm}(0.5\lambda),\hat{\Dm}\big)$}
    
    \addplot [color=red, line width=0.6pt, mark=+, mark options={solid, red}, mark size=2.6pt]
      table[row sep=crcr]{%
	-10	0.00491494873401814\\
	0	0.00156372959480561\\
	10	0.000489524427984945\\
	20	0.000155545512460446\\
	30	4.93883220878325e-05\\
	40	1.55495308794261e-05\\
	50	4.91245381923349e-06\\
	60	1.56629493422986e-06\\
	70	5.5347042757598e-07\\
	80	2.98247250218311e-07\\
    };
    \addlegendentry{RMSE$\big(\bar{\Dm}(0.5\lambda),\hat{\Dm}\big)$}
    
    \addplot [color=red, line width=0.6pt, densely dashed]
      table[row sep=crcr]{%
	-10	0.00491644886022887\\
	0	0.0015547176398062\\
	10	0.000491644886022887\\
	20	0.00015547176398062\\
	30	4.91644886022887e-05\\
	40	1.5547176398062e-05\\
	50	4.91644886022887e-06\\
	60	1.5547176398062e-06\\
	70	4.91644886022887e-07\\
	80	1.5547176398062e-07\\
    };
    \addlegendentry{CRLB$\big(\bar{\Dm}(0.5\lambda)\big)$}

    \addplot [color=blue, line width=0.9pt, dotted, forget plot]
      table[row sep=crcr]{%
    -10	0.000205348979552502\\
    80	0.000205348979552502\\
    };
    \addplot [color=ForestGreen, line width=0.9pt, dotted, forget plot]
      table[row sep=crcr]{%
    -10	2.05297320917998e-06\\
    80	2.05297320917998e-06\\
    };
    \addplot [color=red, line width=0.9pt, dotted, forget plot]
      table[row sep=crcr]{%
    -10	2.54368302726376e-07\\
    80	2.54368302726376e-07\\
    };
    \draw [black] (axis cs:3,8e-6) ellipse [x radius=20, y radius=4];
  	\node[right, align=left]
    at (axis cs:5,7e-6) {\footnotesize{$\sqrt{\mathrm{Tr}(\mathrm{Bias})}$}};
    \end{axis}
    \end{tikzpicture}
    \vspace{-1em}
  \caption{
    Comparison of the derived lower bounds and the \ac{rmse} of the \ac{ml} estimator versus transmit power (SNR).
  }
  \label{Fig_LBRMSE}
\end{figure}

Fig.~\ref{Fig_LBRMSE} presents the comparison of the derived $\mathrm{LB}$ and the tested \ac{rmse} of the \ac{ml} estimator. 
We use $\mathrm{LB}\big(\bar{\Dm}(d),\hat{\Dm}\big)$ to denote the case that generates observations using $\bar{\Dm}(d)$ and estimates using~$\hat{\Dm}$, and the corresponding \ac{rmse} is denoted as RMSE$\big(\bar{\Dm}(d),\hat{\Dm}\big)$.
The classical \ac{crlb} which corresponds to the case $\hat{\Dm}=\bar{\Dm}\big(d)$ is also plotted in Fig.~\ref{Fig_LBRMSE} and denoted as CRLB$\big(\bar{\Dm}(d)\big)$.
In Fig.~\ref{Fig_LBRMSE}, the \acp{rmse} of the \ac{ml} estimator closely follows the derived $\mathrm{LB}$s, demonstrating the validity of our derivation.
The $\mathrm{LB}$s decreases with the increase of $P_\mathrm{T}$ at low-power regions ($\mathrm{MCRB}$ dominates) but finally saturates as $P_\mathrm{T}$ increases ($\mathrm{Bias}$ dominates), which is consistent with the analysis in \textit{Remark~\ref{remark_sat}}. 
Moreover, we notice that a shorter $d$ generates a higher $\sqrt{\mathrm{Tr}(\mathrm{Bias})}$, which indicates a greater performance degradation.
For example, when $d=\unit[0.02]{\lambda}$, the performance starts to saturate at around $P_\mathrm{T}=\unit[10]{dBm}$, which implies that ignoring mutual coupling degrades the channel estimation performance for the regions $P_\mathrm{T}>\unit[10]{dBm}$;
Nonetheless, the mismatch of the case $d=\unit[0.5]{\lambda}$ does not cause a performance saturation until around $P_\mathrm{T}=\unit[70]{dBm}$, which indicates that we can safely ignore the mutual coupling effect in the scenarios $P_\mathrm{T}<\unit[70]{dBm}$ for the setup $d=\unit[0.5]{\lambda}$.

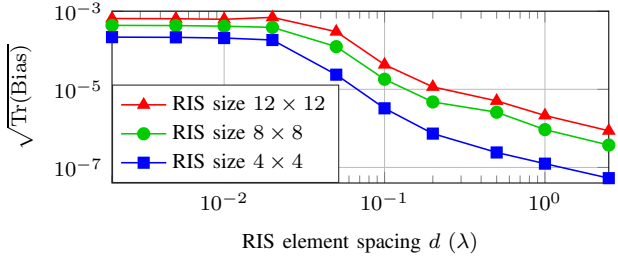
\begin{figure}[t]
    \centering
    \definecolor{ForestGreen}{RGB}{0    204    0}
	\begin{tikzpicture}
	\begin{axis}[%
	width=2.6in, height=0.9in, at={(0,0)}, scale only axis, 
	xmin=0.002,xmax=2.5,xmode=log,xminorticks=true, xlabel={RIS element spacing $d$ $(\lambda)$},
    xlabel style={font=\color{white!15!black},font=\footnotesize},
  	xticklabel style = {font=\color{white!15!black},font=\footnotesize},
	ymin=4e-08,ymax=0.001,ymode=log,yminorticks=true, ylabel={$\sqrt{\mathrm{Tr}(\mathrm{Bias})}$}, 
    ylabel style={font=\color{white!15!black},font=\footnotesize},
  	yticklabel style = {font=\color{white!15!black},font=\footnotesize},
	axis background/.style={fill=white},
	xmajorgrids, ymajorgrids,
	legend style={at={(0,0)}, anchor=south west, legend cell align=left, align=left, font=\footnotesize, draw=white!15!black}
    ]
    \addplot [color=red, mark=triangle*, line width=0.6pt, mark options={solid, red}, mark size=2.4pt]
      table[row sep=crcr]{%
    0.002	0.000649868733334121\\
    0.005	0.000641003490387176\\
    0.01	0.000620114730533513\\
    0.02	0.000694984946932316\\
    0.05	0.00029878450725727\\
    0.1	4.2862430606189e-05\\
    0.2	1.15508409154979e-05\\
    0.5	5.05501157868345e-06\\
    1	2.13950983304466e-06\\
    2.5	8.68650645712992e-07\\
    };
    \addlegendentry{RIS size $12\times 12$}
    
    \addplot [color=ForestGreen, mark=*, line width=0.6pt, mark options={solid, ForestGreen}, mark size=2.2pt]
      table[row sep=crcr]{%
    0.002	0.000433272489555243\\
    0.005	0.000428450083046358\\
    0.01	0.00041258740126188\\
    0.02	0.00038566262961831\\
    0.05	0.000123632424300789\\
    0.1	1.81058373371932e-05\\
    0.2	4.74883773946708e-06\\
    0.5	2.59429520130335e-06\\
    1	9.30195088218431e-07\\
    2.5	3.73969815385829e-07\\
    };
    \addlegendentry{RIS size $8\times 8$}
    
    \addplot [color=blue, mark=square*, line width=0.6pt, mark options={solid, blue}, mark size=2pt]
      table[row sep=crcr]{%
    0.002	0.000216605033790006\\
    0.005	0.000213981851156053\\
    0.01	0.000205311850194109\\
    0.02	0.000183062952736911\\
    0.05	2.37051690298346e-05\\
    0.1	3.26525681126809e-06\\
    0.2	7.35361220036308e-07\\
    0.5	2.41966498713035e-07\\
    1	1.2514300797389e-07\\
    2.5	5.32054209576435e-08\\
    };
    \addlegendentry{RIS size $4\times 4$}
	\end{axis}
	\end{tikzpicture}%
	\vspace{-1.2em}
  \caption{
    Evaluation of $\sqrt{\mathrm{Tr}(\mathrm{Bias})}$ versus \ac{ris} element spacing $d$ for different \ac{ris} size of $\{4\times 4,8\times 8,12\times 12\}$.
  }
  \label{Fig_Biasvsd}
\end{figure}

The above results reveal that the mismatch bias $\sqrt{\mathrm{Tr}(\mathrm{Bias})}$ is a direct metric evaluating the impact of ignoring mutual coupling on channel estimation, since it represents the best performance that can be achieved over all the \acp{snr}.
On the other hand, a lower value of $\sqrt{\mathrm{Tr}(\mathrm{Bias})}$ means it is safe to ignore the mutual coupling effect at a wider region of \ac{snr}.
To manifestly show the impact of ignoring mutual coupling over different \ac{ris} element spacing, Fig.~\ref{Fig_Biasvsd} presents the evaluation of $\sqrt{\mathrm{Tr}(\mathrm{Bias})}$ versus $d$.
In this trial, we test 3 different \ac{ris} sizes as~$\{4\times 4,8\times 8,12\times 12\}$.
The results show that the bias term for all the tested \ac{ris} sizes increases as the \ac{ris} element spacing shortens and finally keeps flat at the region $d<\unit[0.02]{\lambda}$.
This concludes that the shorter the \ac{ris} element spacing, the greater the impact on the considered channel estimation problem until $d<\unit[0.02]{\lambda}$ where the impact keeps flat.
Additionally, it is observed that with a fixed \ac{ris} element spacing, the larger the size of \ac{ris}, the greater the impact of ignoring mutual coupling.

\subsection{Evaluation of Mutual Coupling-Aware Estimation}

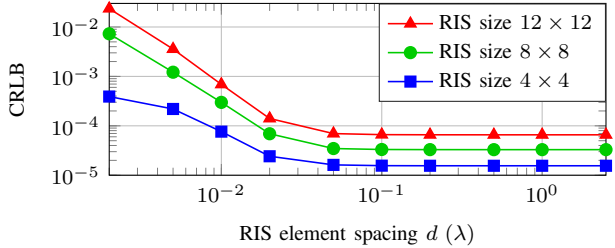
\begin{figure}[t]
    \centering
    \definecolor{ForestGreen}{RGB}{0    204    0}
	\begin{tikzpicture}
	\begin{axis}[%
	width=2.6in, height=0.9in, at={(0,0)}, scale only axis, 
	xmin=0.002,xmax=2.5,xmode=log,xminorticks=true, xlabel={RIS element spacing $d$ $(\lambda)$},
    xlabel style={font=\color{white!15!black},font=\footnotesize},
  	xticklabel style = {font=\color{white!15!black},font=\footnotesize},
	ymin=1e-05,ymax=0.03,ymode=log,yminorticks=true, ylabel={CRLB}, 
    ylabel style={font=\color{white!15!black},font=\footnotesize},
  	yticklabel style = {font=\color{white!15!black},font=\footnotesize},
	axis background/.style={fill=white},
	xmajorgrids, ymajorgrids,
	legend style={at={(1,1)}, anchor=north east, legend cell align=left, align=left, font=\footnotesize, draw=white!15!black}
    ]
    \addplot [color=red, mark=triangle*, line width=0.6pt, mark options={solid, red}, mark size=2.4pt]
      table[row sep=crcr]{%
	0.002	0.0235308099036349\\
	0.005	0.0035736540015027\\
	0.01	0.000689668331535722\\
	0.02	0.000140687155387401\\
	0.05	6.96359102307362e-05\\
	0.1	6.62827547080354e-05\\
	0.2	6.58870936836225e-05\\
	0.5	6.5752554597748e-05\\
	1	6.57655664467365e-05\\
	2.5	6.57656268072601e-05\\
    };
    \addlegendentry{RIS size $12\times 12$}
    
    \addplot [color=ForestGreen, mark=*, line width=0.6pt, mark options={solid, ForestGreen}, mark size=2.2pt]
      table[row sep=crcr]{%
	0.002	0.00729445979425875\\
	0.005	0.0012160064596702\\
	0.01	0.000298223917905634\\
	0.02	6.90995078173411e-05\\
	0.05	3.47059851145356e-05\\
	0.1	3.31796170143793e-05\\
	0.2	3.29880922992771e-05\\
	0.5	3.29361632922402e-05\\
	1	3.29398421601722e-05\\
	2.5	3.29534882377842e-05\\
    };
    \addlegendentry{RIS size $8\times 8$}
    
    \addplot [color=blue, mark=square*, line width=0.6pt, mark options={solid, blue}, mark size=2pt]
      table[row sep=crcr]{%
	0.002	0.000391029355970992\\
	0.005	0.000219190605885329\\
	0.01	7.62717686062655e-05\\
	0.02	2.4212193281156e-05\\
	0.05	1.62025600252016e-05\\
	0.1	1.56266730324487e-05\\
	0.2	1.5555483460606e-05\\
	0.5	1.55441489065255e-05\\
	1	1.55490638322639e-05\\
	2.5	1.55507939147246e-05\\
    };
    \addlegendentry{RIS size $4\times 4$}
	\end{axis}
	\end{tikzpicture}%
	\vspace{-1.2em}
  \caption{
    Evaluation of the mismatch-free \ac{crlb} versus \ac{ris} element spacing $d$ for different \ac{ris} size of $\{4\times 4,8\times 8,12\times 12\}$ with $P_\mathrm{T}=\unit[40]{dBm}$.
  }
  \label{Fig_CRLBvsd}
  \vspace{-0.5em}
\end{figure}

Now we evaluate the impact of the mutual coupling effect in the mutual coupling-aware channel estimation.
Suppose the estimator has the information of $\Zm_\mathrm{SS}^{\mathrm{mutual}}$, i.e., $\hat{\Dm}=\bar{\Dm}$, thus the estimation \ac{rmse} is lower bounded by the classical \ac{crlb}, as clarified in \textit{Remark~\ref{remark_4}}.
Fig.~\ref{Fig_CRLBvsd} shows the evaluation of the mismatch-free \ac{crlb} versus \ac{ris} element spacing $d$ for different \ac{ris} size of $\{4\times 4,8\times 8,12\times 12\}$, where the transmit power is fixed as $P_\mathrm{T}=\unit[40]{dBm}$.
It unveils that at the large-spacing regions ($d>\unit[0.05]{\lambda}$), mutual coupling does not affect the performance bounds in a mutual coupling-aware estimation.
However, when the \ac{ris} element spacing $d<\unit[0.05]{\lambda}$, the CRLBs increase with the decrease of $d$.
This shows that even with the mismatch-free setup (i.e., the mutual impedance $\Zm_\mathrm{SS}^{\mathrm{mutual}}$ is known), the mutual coupling effect can degrade the channel estimation performance if the \ac{ris} element spacing is too small ($d<\unit[0.05]{\lambda}$).
Finally, we have the observation that for a fixed $d$, the larger the size of \ac{ris}, the higher the \ac{crlb}.
This is because a larger RIS size increases the dimension of the unknown channel vector, making channel estimation more challenging.

\textit{Lessons Learned:} 
As the densification of RIS unit cells amplifies the impact of mutual coupling, particularly in applications like holographic MIMO~\cite{An2023Tutorial}, disregarding this effect becomes untenable. By carefully adjusting the RIS configuration, it is possible to mitigate the influence of mutual coupling, as demonstrated in~\cite{Qian2021Mutual}, thereby enhancing channel estimation performance. Nevertheless, such an optimization over RIS configuration necessitates knowledge of mutual impedance among RIS unit cells, which can be obtained through a calibration process, either online or offline. Practically, the RIS configuration and channel estimation can be performed alternately, which may augment pilot transmission, indicating a trade-off between pilot overhead and estimation performance.

\section{Conclusion}
This study examined the impact of mutual coupling on channel estimation within a \ac{ris}-assisted communication system. 
Specifically, we derived the \ac{mcrb} of \ac{ris}-assisted channel estimation, which accounts for the model mismatch caused by the neglect of mutual coupling. 
Our analysis and numerical results reveal that ignoring mutual coupling in scenarios involving closer integration of RIS elements or larger RIS size leads to a more pronounced degradation in channel estimation performance.
Conversely, even with mutual coupling-aware setups, too tight RIS element spacing can also lead to significant estimation performance degradation. 
Future work can focus on the development of mutual coupling-aware channel estimation algorithms, antenna array design, and improvement of the computing ability of metasurfaces.

\bibliography{references}
\bibliographystyle{IEEEtran}

\end{document}